\documentclass[9pt]{article}  \usepackage{times}
\usepackage{graphicx}

\usepackage{color}

\usepackage[round,numbers,sort&compress]{natbib} 
\usepackage{amsmath}
\usepackage{amssymb}
\usepackage{stackrel}
\usepackage{chemarrow}
\usepackage{graphicx}
\usepackage{textgreek}
\usepackage{booktabs}
\usepackage{dcolumn}
\usepackage{rotating}
\usepackage{multirow}
\renewcommand\appendix{\par
  \setcounter{section}{0}
  \setcounter{subsection}{0}
  \setcounter{figure}{0}
  \setcounter{table}{0}
  \renewcommand\thesection{Appendix \Alph{section}}
  \renewcommand\thefigure{\Alph{section}\arabic{figure}}
  \renewcommand\thetable{\Alph{section}\arabic{table}}
}

\begin{document}

%--- the below commands make the spacing around new sections smaller
%\titlespacing{\section}{0pt}{2pt}{0pt}
%\newenvironment{sciabstract} {\begin{quote} \bf} {\end{quote}}
\twocolumn[{\LARGE \textbf{Solitary electromechanical pulses in Lobster neurons.\\*[0.2cm]}}
{\large Alfredo Gonzalez-Perez$^1$, Lars D. Mosgaard$^1$, Rima Budvytyte$^1$, Edgar Villagran-Vargas$^3$, Andrew D. Jackson$^2$ and Thomas Heimburg$^{1,\ast}$\\*[0.1cm]
{\small $^1$Niels Bohr Institute, University of Copenhagen, Blegdamsvej 17, 2100 Copenhagen \O, Denmark}\\*[-0.1cm]
{\small $^2$Niels Bohr International Academy, University of Copenhagen, Blegdamsvej 17, 2100 Copenhagen, Denmark}\\*[-0.1cm]
{\small $^3$Facultad de Ciencias, Universidad Aut\'onoma del Estado de M\'exico. Instituto Literario 100, Colonia Centro, 50000, Toluca, M\'exico}\\

{\normalsize \textbf{ABSTRACT}\hspace{0.5cm} Investigations of nerve activity have focused predominantly on electrical phenomena. Nerves, however, are thermodynamic systems, and changes in temperature and in the dimensions of the nerve can also be observed during the action potential.  Measurements of heat changes during the action potential suggest that the nerve pulse shares many characteristics with an adiabatic pulse. First experiments in the 1980s suggested small changes in nerve thickness and length during the action potential. Such findings have led to the suggestion that the action potential may be related to electromechanical solitons traveling without dissipation. However, they have been no modern attempts to study mechanical phenomena in nerves. Here, we present ultrasensitive AFM recordings of mechanical changes on the order of 2 -- 12 {\AA} in the giant axons of the lobster. We show that the nerve thickness changes in phase with voltage change. When stimulated at opposite ends of the same axon, colliding action potentials pass through one another and do not annihilate. These observations are consistent with a mechanical interpretation of the nervous impulse. 
\\*[0.3cm] }}
\noindent\footnotesize{\textbf{Keywords:} action potential, atomic force microscopy, nerve pulse collision, heat, capacitance\\*[0.1cm]}
\noindent\footnotesize {$^{\ast}$corresponding author, theimbu@nbi.dk. }\\
\vspace{0.3cm}
]
%------------------------------------------------------------------

%\noindent\footnotesize {$^{\ast}$corresponding author, theimbu@nbi.dk. }\\

%\noindent\footnotesize{\textbf{Keywords:} Nerves, thermodynamics, hydrodynamics, atomic force microscopy, electrophysiology, solitons}\\
%\noindent\footnotesize{\textbf{Abbreviations:} 

\normalsize
%====================================================
\section*{Introduction}
It is not generally appreciated that the size of the nervous impulse is remarkably large. A myelinated motor neuron has a pulse velocity on the order of 100 m/s. Given a typical pulse duration of 1 ms, the resulting size of the nerve pulse is 10 cm.  Propagation velocities in slow, non-myelinated fibers are reduced to  1-10 m/s, indicating a pulse size of 1-10 mm. Thus, nerve pulses are macroscopic phenomena spanning a significant fraction of the total axon length. In some cases they can even be larger than small neurons such as interneurons that are only a few 100 $\mu$m long \cite{Goodman1974}.  In the past, the activity of nerves has conventionally been considered to be a purely electrical phenomenon produced by the flux of ions and the charging of the membrane capacitor \cite{Hodgkin1952b}. The shear size of a nerve pulse suggests that the macroscopic thermodynamic properties of the nerve membrane or of the entire nerve ought to be taken into account.  It is to be expected that the state of the nerve cell depends not only on electrochemical potentials and the conjugated flux of ions but also on all other thermodynamic forces including variations in lateral pressure (resulting in changes of membrane area and thickness) and temperature (resulting in heat flux). It is therefore not surprising that, during the action potential, one finds changes not only in voltage but also in thickness \cite{Iwasa1980a, Iwasa1980b, Tasaki1989, Kim2007}, length \cite{Wilke1912b, Tasaki1989} as well changes in membrane temperature \cite{Abbott1958, Howarth1968, Howarth1975, Ritchie1985}. The change of thickness of a single squid axon was found to be on the order of 1 nm, and temperature changes range between 1--100 $\mu$K depending on the specimen. While both mechanical and thermal signals are very small, they are found to be in phase with voltage changes. The heat signal can be blocked by neurotoxins such as tetrodotoxin \cite{Ritchie1985}. This strongly suggests that these thermodynamic phenomena are correlated with the voltage pulse and do not represent independent secondary phenomena.

From a thermodynamic viewpoint, one can consider two extreme cases of possible dynamic changes associated with the action potential: 1. Purely dissipative processes during which entropy increases, such as the flow of ions along concentration gradients. Such processes form the basis of the Hodgkin-Huxley model for the action potential.  2. Adiabatic processes that do not dissipate heat and thus conserve entropy. These phenomena are rather governed by the laws of analytical mechanics. They play an important role for dynamic properties such as sound propagation. In this context, the heat changes observed in nerves are of fundamental interest. It was found that heat is released during an initial phase of the action potential and that this heat is reabsorbed in the final phase of the action potential.  During the nerve pulse no heat (or only very little) is dissipated, and the entropy of the membrane is basically conserved \cite{Howarth1968, Ritchie1985}. Thus, thermal measurements suggest that the action potential is an adiabatic phenomenon reminiscent of a sound wave. As early as 1912, the striking absence of heat production led Hill to conclude: \textit{`This suggests very strongly ... that the propagated nervous impulse is not a wave of irreversible chemical breakdown, but a reversible change of a purely physical nature'} \cite{Hill1912}. In contrast, the contemporary understanding of the nerve pulse is based on the flow of ions along gradients through ion channel proteins \cite{Hodgkin1952b} and therefore assumes that it is of a dissipative nature. Hodgkin himself compared it to the \textit{`burning of a fuse of gunpowder'} \cite{Hodgkin1964}.  There is thus a disagreement between electrophysiological models and some thermodynamics findings. These problems are not easily resolved and merit careful attention. 

\begin{figure}[htb!]
    \centering
	\includegraphics[width=8.5cm]{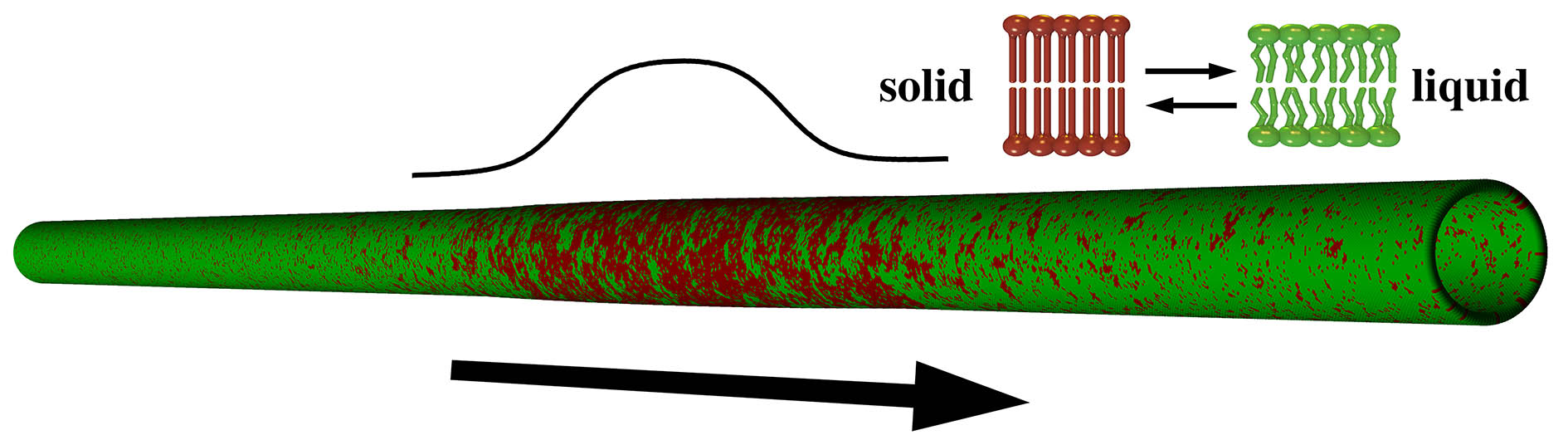}
	\parbox[c]{8 cm}{ \caption{\textit{Schematic representation of a mechanical soliton traveling in a nerve fiber (according to \cite{Heimburg2005c}). Small local changes in thickness are caused by pressure-induced order transitions in the membrane. Red regions correspond to ordered lipids in the otherwise liquid (green) lipid membrane. In a living nerve, the spatial extension of the pulse is much larger than shown here.}
	\label{Figure1}}}
\end{figure}

The observed reversible change in temperature as well as mechanical changes seen in optical and mechanical experiments led to the proposal that the action potential is a consequence of an electromechanical pulse or soliton \cite{Heimburg2005c}.  A condition for the existence of such a soliton is the existence of an order transition in the membrane from solid to liquid slightly below physiological temperature.  Such transitions have been found in various biological membranes \cite{Heimburg2007a}. It is thought that the soliton consists of a region of ordered lipid membrane traveling in the otherwise liquid membrane with a speed somewhat less than the speed of sound in the membrane \cite{Heimburg2005c, Lautrup2011}. This is shown schematically in Fig. \ref{Figure1}. The difference in membrane thickness between the solid and the liquid is of order 1 nm ($\approx$17\% of the total membrane thickness). The associated reversible change in energy is related to the latent heat of the membrane transition and thus to the reversible heat production found in nerves. Since the membrane changes its thickness, changes in membrane voltage of order 50mV are to be expected as a consequence of changes in its capacitance \cite{Heimburg2007b, Heimburg2012}. Thus, the soliton is of an electromechanical or piezoelectric nature. An intrinsic feature of such solitons is that two colliding pulses pass through each other without dissipation \cite{Lautrup2011} rather than annihilating as expected in Hodgkin-Huxley-like pulses due to the refractory period. In fact, the penetration of colliding nerve pulses was seen recently in nerves from earth worm and the ventral cord of lobster \cite{GonzalezPerez2014}. The soliton model treats the pulse as a longitudinal compressional density change that is strongly affected by the presence of a phase transition in the membrane. The velocity of the pulse is closely related to the sound velocity in a liquid lipid membrane. It predicts pulse velocities close to 100 m/s, very similar to those in myelinated nerves. In the soliton model, the pulse velocity is not related to the axon radius, while it depends on the square root of the radius in the HH-model. However, there is evidence that the radius dependence in real nerves deviates from the latter behavior. For instance, Goldman \cite{Goldman1964} found that the 4-fold stretching of a single neuron (equivalent to a 4-fold decrease in radius) did not lower the pulse velocity. In fact, upon stretching the pulse velocity first increased and then stayed constant over a significant range of axon radii.
%vvvvvvvvvvvvvvvvvvvvvvvvvvvvvvvvvvv
%^^^^^^^^^^^^^^^^^^^^^^^^^^^^^^^^^^^

Changes in voltage, thickness and temperature in a soliton are all associated with a single phenomenon, and it is not appropriate to consider some of these changes as side effects of another dominant process. They are rather different aspects of the same phenomenon as seen by different instrumentation. There exists clear evidence that electromechanical pulses can travel on lipid monolayers close to the LE-LC transition with velocities very close to those of non-myelinated nerves \cite{Griesbauer2012a, Griesbauer2012b}. However, there is a striking lack of experiments that actually demonstrate the mechanical nature of the nerve pulse. For this reason, the thermodynamic and mechanical interpretation of the nerve pulse has been widely ignored.

%====================================================
\section*{Materials and methods}

\emph{\textbf{Materials.}} In our experiments we used lobster, \textit{Homarus americanus}, that was obtained from a local supplier that imported the animals from Canada. We used a lobster saline solution adapted from Evans et al.  \cite{Evans1976} 462 mM NaCl, 10 mM KCl, 25 mM CaCl$_2$, 8 mM MgCl$_2$, 10 mM TRIS and 11 mM Glucose, adjusted to pH 7.4 with NaOH.  All chemicals used in the preparation were purchased from Sigma-Aldrich.\\*[0.1cm]
\noindent\emph{\textbf{Recording of the electromechanical action potential.}}
The electrical signal was recorded using a Powerlab 8/35 (ADInstruments Europe, Oxford, UK). It was preamplified with a differential amplifier DP-304 from Warner Instrument Corporation with a gain x1000 using a 3 kHz lowpass and a 10 Hz  highpass filter. The mechanical displacement was obtained using an AFM, NanoWizard II from JPK (Germany). The signal was fed into the PowerLab 8/35 and analyzed simultaneously with the electrical recording without any preamplification. 
The AFM was mounted on the top of an inverted microscope Olympus IX71 (Olympus Corporation, Japan) placed on an anti-vibration table TS-150 (low power, from Herzan LLC, USA). The full setup containing the AFM, optical microscope as well as the anti-vibration table was placed inside a Faraday cage in order to avoid external electrical noise. We used tipless cantilevers from  Mikromasch Europe (type HQ:CSC37 and HQ:CSC38). The resonance frequency is 20 kHz  for the HQ:CSC37 cantilever and the force constant is 0.3 N/m. The resonance frequency is 10 kHz for the HQ:{\-}CSC38 cantilever and the force constant is 0.03 N/m. Thus, the resonance frequencies are outside of the range of the expected signal (1-2 \textmu s). Both cantilevers were used under the same experimental conditions. The recording signal frequency was 40 kHz for both, the electrical and mechanical measurement. All the experiments were performed at room temperature of about 22 C.
%vvvvvvvvvvvvvvvvvvvvvvvvvvvvvvvvvvv
\begin{figure*}[htb!]
    \centering
	\includegraphics[width=14cm]{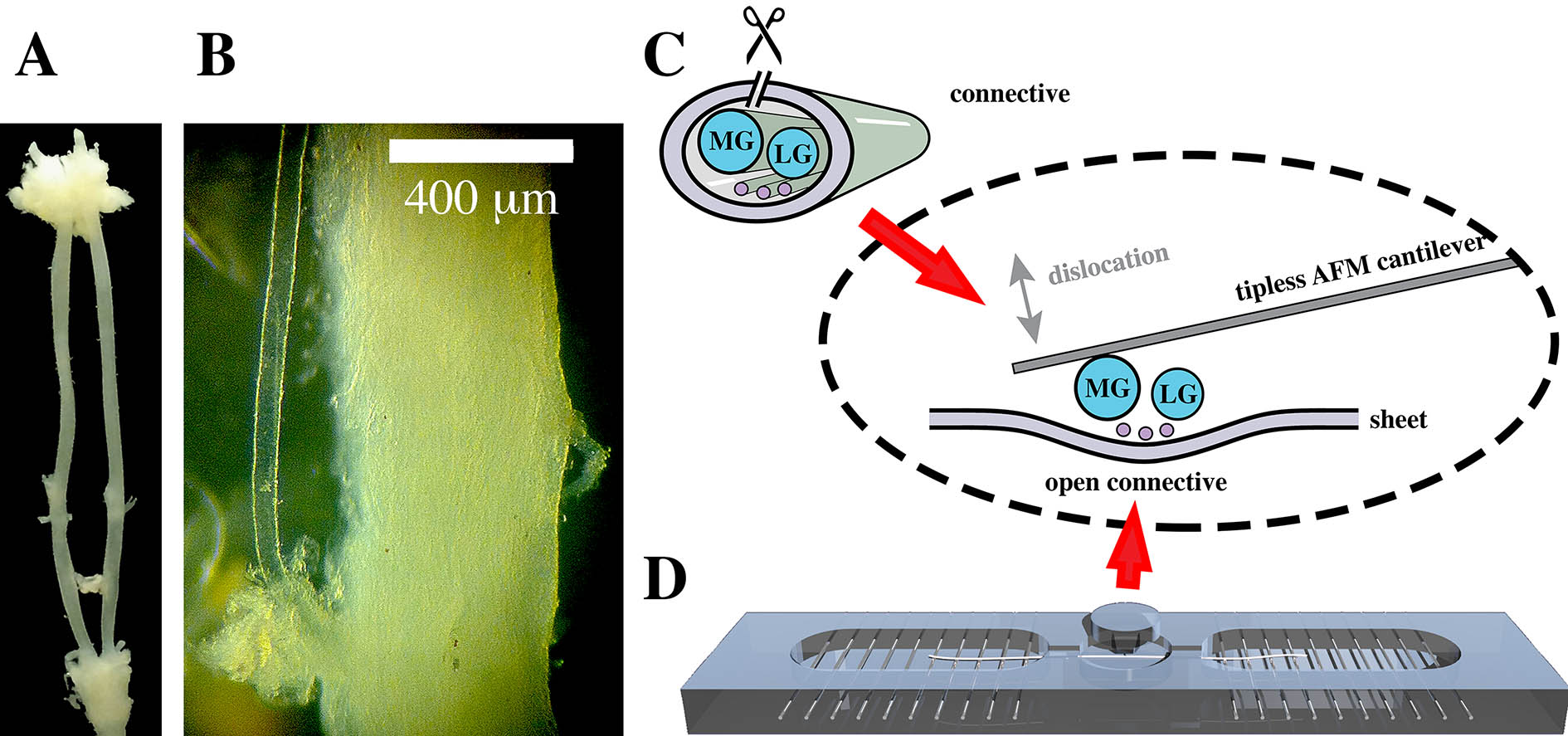}
	\parbox[c]{15cm}{ \caption{\textit{Outline of the mechanical experiment. A. Image of the two lobster connectives including the brain (top) and the subesophageal ganglion (bottom). B. Detail of a lobster connective with the lateral giant axon exposed. C.  Schematic representation of the cross-section of a lobster connective containing a medial giant axon, a lateral giant axon and several small fibers. The sheath surrounding the connective is cut in the longitudinal direction. D. After opening the sheats, the two giant axons are exposed to the outside solution. Mechanical signals are recorded with a tipless AFM cantilever. The open connective is placed on a cell containing several pairs of electrodes to stimulate the nerve and to record the signal. The mechanical response is measured on a planar support in the middle of the cell.}
	\label{Figure2}}}
\end{figure*}\\*[0.1cm]
\noindent\emph{\textbf{Nerve chamber.}} We used two different nerve chambers. For the collision experiment the nerve chamber is composed by an array of 21 stainless steel electrodes in a longitudinal cavity covered by a lid in order to isolate the nerve once extracted. The lid also allows to keep the nerve in a saturated vapor atmosphere to prevent the ventral cord from drying. The nerve chamber is a 7 x 2.5 cm block by 1 cm height made on Plexiglass that contains two longitudinal perforation of 1.5 cm length by 0.5 cm wide and 0.5 cm width \cite{GonzalezPerez2014}. The large longitudinal aperture contains an array of 21 perforations to allocate the stainless steel electrodes. The array was placed about 0.25 cm from the top of the chamber. The distance between two consecutive electrodes is of 0.25 cm. The stainless steel electrodes have a length of about 3.4 cm and a diameter of 0.5 mm and were fixed in the perforation along the chamber by using a rubber replica casting system (Reprorubber, Islandia, NY) composed by a base and a catalyst from Flexbar Machine Corp (Islandia, NY). For the AFM experiment we used a variation of the previous chamber with half size thickness and with a small open area in the middle without electrodes that allows access to the AFM cantilever. \\*[0.1cm]
\noindent\emph{\textbf{Sample preparation.}} The Lobster, \textit{Homarus americanus}, was anesthetized by keeping the animal in the freezer for about 30 min. Once removed from the freezer the animal was placed on the dissecting table and the legs were removed. A second cut was made at the beginning of the abdomen to separate the abdominal part from the thorax. In the thorax, we removed the carapace and made the extraction dorsally. The ventral cord is contained in a tube-like structure formed by the exoskeleton of the lobster and is easy to remove after breaking the shell with scissors.  The thoracic segment of the ventral cord including the brain was extracted as a single piece. Special care was taken to remove the circumesophageal connectives from their location surrounding the esophagus. The brain, circumesophageal connectives and subesophageal ganglia was removed from the rest of the ventral cord as placed in a petri-dish with lobster saline solution. The basic anatomical features as well as the basic steps in the preparation are described in the literature  \cite{Wright1958}.

The two circumesophageal connectives were severed at the level of the brain and the subesophageal ganglia (Fig. 2A). From each circumesophageal connective we remove the external connective sheath to expose the main median giant axon (with diameters between 150 and 200 $ \mu$m, cf. Fig. 2B and C). The preparation was transferred to the AFM nerve chamber with the medial giant axon facing the AFM cantilever (Fig. 2D). For the collision experiment the external connective layer was not removed ensuring a longer survival of the nerve. One of the giant axons was cut at one or two position to rule out the possibility that propagation of pulses in opposite directions occurs in different axons. The total length of the connectives used by us was $\approx$ 4-5 cm.\\*[0.1cm]
\noindent\emph{\textbf{Collision experiments.}}
The collision experiment was performed independently using a PowerLab 26T from ADInstruments. The instrument possesses an internal bio-amplifier that allows the recording of small electrical potential on the order of microvolts. The bio-amplifier contains two recording channels (further description see ADInstruments webpage). We used the Labchart software from ADInstruments in order to record the signals coming from the ventral cord. The recording frequency was 40 kHz.  All experiments were performed at room temperature of about 22 C.

In some experiments we cut the LG axon of the connective at two locations in order to observe a collision of pulses in the MG axon only (cf. Fig. 4, right). In order to make sure that the same axon was cut twice, we performed the following tests: \\
1. The intact connective (before removing the sheath and cutting the LG axon), we observed two signals from each side corresponding to the action potentials in the MG and LG axons. 
2. We performed on single cut at the one end of LG axon. Then, connective was tested again by stimulating at one end of axon before the cut and recording after cut. Only a single peak originating from the MG axon was observed. When placing the stimulation electrode behind the cut, two peaks from the LG and MG axons could be observed.
3. A second cut was made close to the other end of connective. When placing the stimulation electrodes before the first cut and the recording electrodes after the second cut, only one action potential originating from the MG axon was observed, while LG signal was absent. \\*[0.1cm]
\noindent\emph{\textbf{Ethical.}} The work described in this article has been carried out in accordance with the policy on the use of animals of the Society for Neuroscience.
%vvvvvvvvvvvvvvvvvvvvvvvvvvvvvvvvvvv
\begin{figure*}[htb!]
    \centering
	\includegraphics[width=14cm]{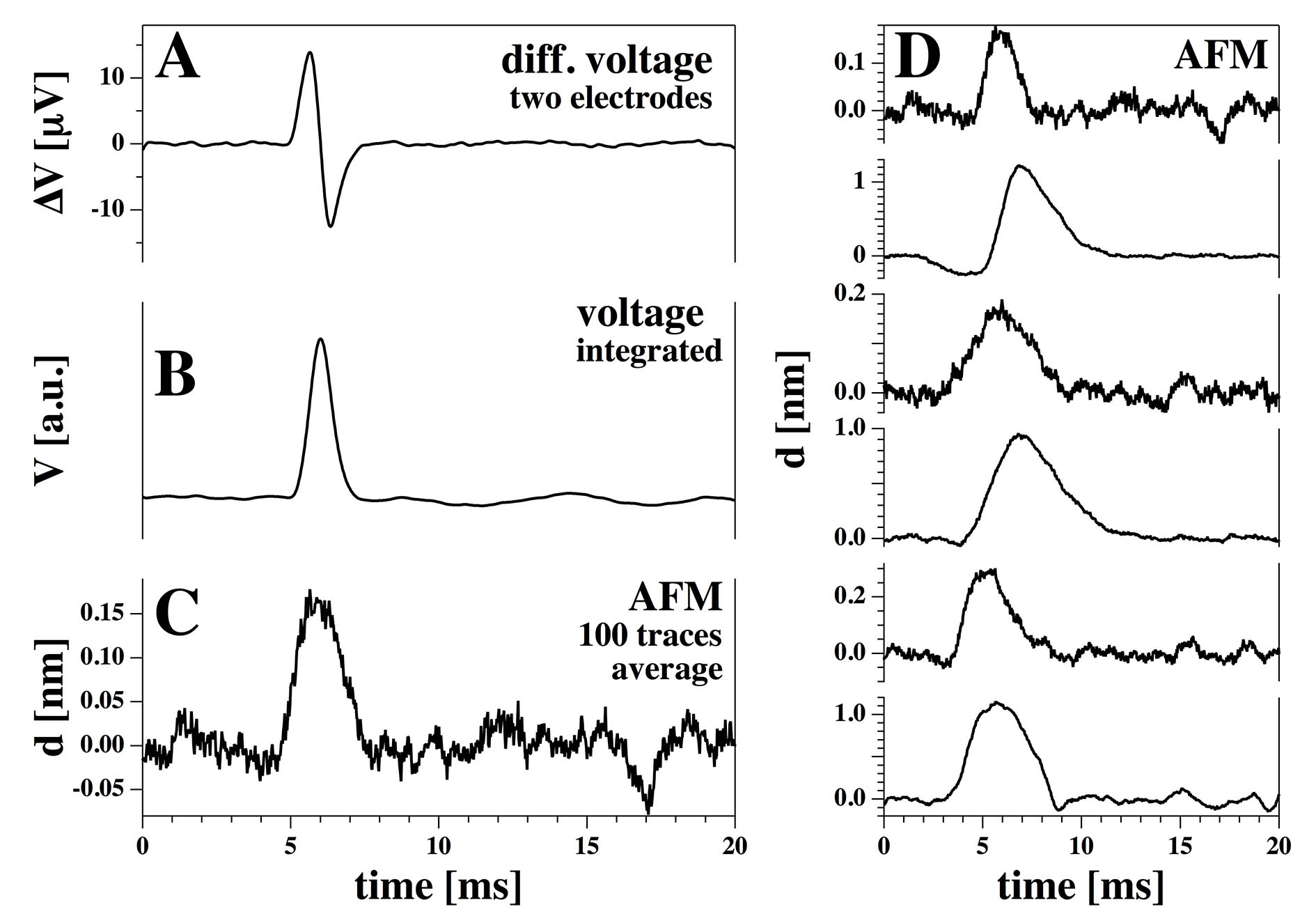}
	\parbox[c]{15cm}{ \caption{\textit{Vertical displacements as recorded by AFM. Left: Differential voltage change (A), integrated signal (B) and corresponding mechanical displacement in the same neuron (C). Right: Mechanical recordings from six different preparations all yield mechanical changes between 0.2 and 1.2 nm. All recordings have a similar shape with small differences due to the size of the nerve and the precise positioning of the AFM cantilever. Mechanical and electrical signals were not measured at exactly the same location. In this figure, they have been temporally aligned.}
	\label{Figure3}}}
\end{figure*}
%^^^^^^^^^^^^^^^^^^^^^^^^^^^^^^^^^^^
%vvvvvvvvvvvvvvvvvvvvvvvvvvvvvvvvvvv
\begin{figure*}[htb!]
    \centering
	\includegraphics[width=14cm]{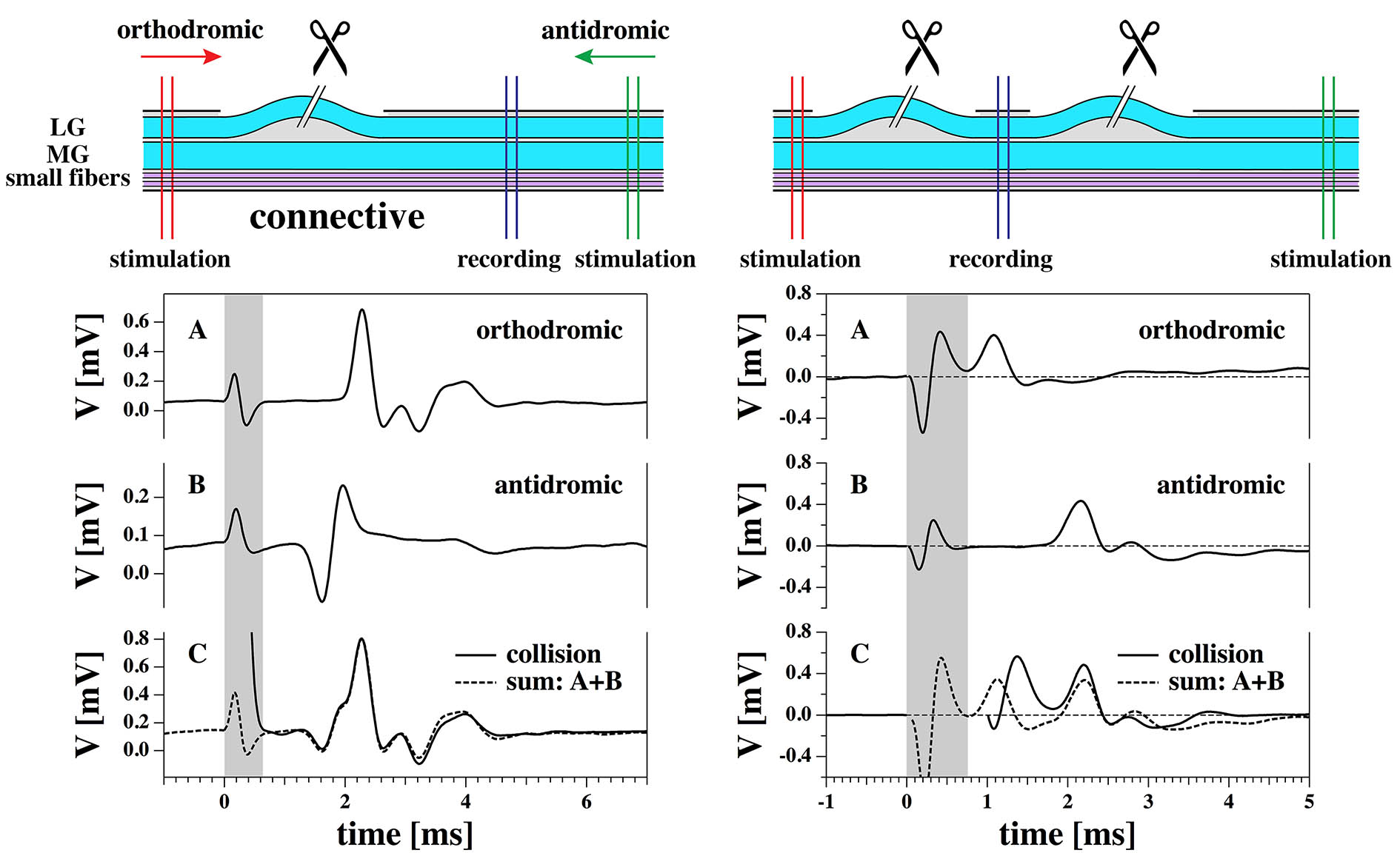}
	\parbox[c]{15cm}{ \caption{\textit{Pulse collision experiment in the median axon of the lobster connective. The top panels represent a schematic drawing of the connective, the position of the electrodes and of the cuts in the lateral giant axon. Left: The lateral giant axon was cut at one position on the orthodromic side of the recording electrodes. Trace A. The nerve was stimulated in orthodromic direction with the two red electrodes  The signal was recorded with the two blue electrodes. Trace B. The nerve was stimulated in antidromic direction with the two green electrodes and recoded with the blue electrodes.Trace C. The nerve was stimulated from both sides. Both pulses can be measured with the blue electrodes, suggesting that they did not annihilate upon collision. The dashed line is the sum of the orthodromic and the antidromic signals from traces A and B and is given for comparison. Right: A similar experiment in which the lateral giant axon was cut at two positions left and right of the recording electrodes. In this experiment, both orthodromic and antidromic pulses could be seen at the recording electrodes suggesting that the did not annihilate upon collision. The grey-shaded regions display the stimulation artifact. For experimental details see also \cite{GonzalezPerez2014}.}
	\label{Figure4}}}
\end{figure*}
%^^^^^^^^^^^^^^^^^^^^^^^^^^^^^^^^^^^%----- begin RESULTS ----------------------------------------------------------
\section*{Results}
\subsection*{Mechanical changes}
Here, we present the results of experiments on the mechanical changes in single axons from lobster connectives. %(Fig. \ref{Figure2}A and B). 
The ventral cord of the lobster is made of two connective strands containing one lateral and one median giant axons each (LG and MG axons, respectively). Additionally, they contain several small nerve fibers. The action potentials of the giant axons can be clearly distinguished from those of the small fibers. The latter yield much smaller signals and are excited only at higher voltages. The MG axon displays a larger peak amplitude and a larger conduction velocity as compared to the LG axon.

We recorded the mechanical response of the giant axons using atomic force microscopy (AFM). Since the vertical dislocation of the axonal membrane is expected to be small, the surface of individual giant axons must be accessed directly by the AFM cantilever. In order to achieve this, the sheath surrounding the connective was cut open. This provides direct access to the single axons (a single exposed LG axon is shown in Fig. \ref{Figure2} B together with the opened connective). The experimental setup is described schematically in Figs. \ref{Figure2}C and D, and described in detail in the Materials and Methods section.  In brief, a tipless cantilever is placed on a single axon exposed from the opened connective. The open connective is placed on top of a cell with 21 electrodes that allow us to stimulate the nerve and to measure the electrical response. In the center of the cell, the nerve is placed on a support for the mechanical measurement. We stimulated the nerve periodically with a pair of electrodes at one end. We monitor only the vertical displacement of the cantilever without scanning in the $x$-$y$ plane. The sampling rate is 40 kHz, and the response time of the AFM setup is about 1 ms. The experimental results are shown in Fig. \ref{Figure3} (left panels). The nerve signal is recorded by two electrodes separated by  a distance less than the width of the nerve pulse. Thus, in effect, the first derivative of the true voltage change is measured (Fig. \ref{Figure3}A). The integral of this signal is roughly proportional to the true voltage change (Fig. \ref{Figure3}B). The bottom trace is the AFM signal (Fig. \ref{Figure3}C). In order to achieve a good signal-to-noise ratio we averaged over approximately 100 individual action potentials, excited at intervals of 0.2 seconds. In the experiment shown in Fig. \ref{Figure3}C, the vertical displacement was about 2 {\AA}. Fig. \ref{Figure3}D) shows mechanical signals from six different lobster specimens. We find displacements between 2 and 12 {\AA} lasting between $\sim$2 and 4 ms. Thus, the mechanical changes in these neurons can be measured in a consistent and reproducible manner. We found these mechanical signals consistently in all connective preparations studied. Electrical recordings demonstrated that more than one single axon was stimulated in the recordings with larger voltage amplitudes. The stimulation artifact visible in the electrical recordings had no influence on the mechanical recording. This suggests that the measured displacement is not due to a direct influence of voltage on the cantilever.

All of our recordings were averaged over $\approx$ 100 pulses. Our measurements displayed some variance of the thickness change between 0.15-1.2 nm, which is partially related to the quality of the contact between the cantilever and the neuronal membrane.Therefore, the displacement of 0.15 nm as in Fig. \ref{Figure3}C should be regarded as a lower limit of the true dislocation. Further, the mechanical amplitude depends on how many neurons fire at the same time. Most recordings displayed a displacement around 1 nm (partially shown in Fig. \ref{Figure3}D). A thickness change of 1nm is very significant because it is close to the thickness difference between a liquid and a solid membrane with the correct sign. This is exactly the change expected in the electromechanical soliton model. \\

It is important to notice that we find that the mechanical changes are proportional to the voltage changes (see Fig. \ref{Figure3} B and C). This is consistent with an electromechanical interpretation of the coupling between voltage and membrane thickness as put forward by \cite{Heimburg2012} and \cite{ElHady2015}. However, this result deviates from the data by Iwasa and Tasaki \cite{Iwasa1980a, Iwasa1980b} who find that the mechanical trace resembles the first derivative of the voltage trace as it is typically obtained in extracellular recordings with two electrodes on the neuron (personal communication with K. Iwasa, cf. Fig. \ref{Figure3}A). In this respect, our data are not merely a confirmation of previous data but actually provide new evidence. This will be discussed in more detail in the Discussion section.\\

\subsection*{Pulse collision}
%It is clear that both electrical and mechanical phenomena exist in various nerves. 
In a recent publication we showed that action potentials traveling in opposite direction in some nerves (earth worm, lobster ventral cords) can pass through each other upon collision \cite{GonzalezPerez2014}. Whether this is a generic feature of nerves is not clear. In a very early experiment Tasaki showed that colliding pulses in single axons from sciatic nerves of toads annihilate \cite{Tasaki1949}. Penetration of pulses speak in favor of a mechanical mechanism for nerve pulse propagation. In contrast, annihilation favors a dissipative ion-channel based view. 

Here, we present evidence for pulse penetration in single axons from the connectives that were used in the mechanical experiment. We stimulated an action potential in orthodromic and in antidromic directions. We performed experiments on about 20 lobsters. The voltage was chosen such that only one axon was exited. In about 6 experiments, we severed one axon of the connective in order to eliminate the possibility that pulses in opposite directions traveled in different axons (indicated in Fig. \ref{Figure4}, top left). In an additional 3 experiments, we cut one axon at two different locations (indicated in Fig. \ref{Figure4}, top right, see Materials and Methods section for details). In these experiments, there exists only one giant axon between the three pairs of electrodes.  Stimulation was made using pairs of electrodes at the ends of the axon.  Propagating pulses were recorded with a pair of electrodes at a position closer to the antidromic side (indicated in Fig. \ref{Figure4}, top left) or closer to the orthodromic side (indicated in Fig. \ref{Figure4}, top right). The intact giant axon (the MG fiber) was stimulated in the orthodromic direction (trace A), in the antidromic direction (trace B) and in both directions (trace C). When increasing the voltage it became evident that only one giant axon could be stimulated. Some residual activity from the small fibers could sometimes be seen. In the experiment shown in Fig. \ref{Figure4}, left), the antidromic signal arrives at the recording site prior to the orthodromic signal. When stimulated at both ends, the resulting trace looked similar to the sum of single orthodromic and single antidromic pulses indicating that the pulses penetrated without major perturbation. This is consistent with previous findings in worm axons and other axons in lobster \cite{GonzalezPerez2014}. In the experiment shown in Fig. \ref{Figure4}, right, the antidromic signal arrives at the recording site after the orthodromic signal. After collision, both pulses can still be seen, indicating that they passed through each other. We have shown that the same events can also be seen when nerves are inserted into capillaries and the nerves are complete surrounded by saline solution (data not shown).

%----- end RESULTS ------------------------------------------------------------
%----- begin DISCUSSION ------------------------------------------------------
\section*{Discussion}
Here, we have reported mechanical changes in single axons from lobster connectives. These are the first AFM recordings of an action potential in a single axon, and they are probably the most sensitive recordings of mechanical changes in a single axon so far. We have also reported collision experiments in the single axons that show that action potentials traveling from the two ends of an axon can pass through each other without being annihilated. We discuss below why such findings are important.

Mechanical recordings of nerve pulses were pioneered by Iwasa \& Tasaki in the early 1980s \cite{Iwasa1980a, Iwasa1980b}. They reported displacements of the membrane of the squid axon on  the order of 1 nm. While the magnitude of the displacement in our experiments is of the same order, the functional form of their displacement differs from ours. In our data the displacement is proportional to the voltage change in the axon, whereas the data by Iwasa \& Tasaki are in fact proportional to the first derivative of the voltage. Thus, in their experiments the mechanical trace resembles the one shown in Fig. \ref{Figure3}A, while our data resemble the traces shown in Fig. \ref{Figure3}B. Our results are consistent with data by Kim et al. \cite{Kim2007} on synapse bundles, where the mechanical changes were also found to be proportional to the voltage change. Optical recordings of dimensional changes in nerves also indicate that displacements are proportional to voltage \cite{Salzberg1989, Kim2007}. This difference in functional form is important because Iwasa's \& Tasaki's data are not consistent with the concept of electrostriction, while the data presented here are actually in good agreement with this concept (see Appendix A for details). For this reason, the data by Iwasa \& Tasaki are also not consistent with the soliton model put forward in \cite{Heimburg2005c}. The origin of the deviation of the early mechanical data from other recordings is not quite clear. Later direct recordings of mechanical changes induced by voltage include \cite{Zhang2001}, who determined movement of membranes induced by voltage changes across HEK cell membranes by AFM (they find a displacement proportional to voltage), and \cite{Nguyen2012}, who measured mechanical changes in the soma of rat PC12 cells with a piezo-electric ribbon (reporting a force generated by a membrane proportional to voltage). The latter publication pointed out that AFM experiments on single neurons are in fact very difficult. These reports justify the need of studies such as presented here.

Our experimental findings are embedded into a discussion about the validity of the Hodgkin-Huxley model, and the possibility of the existence of thermodynamic phenomena. Both the mechanical changes reported here, and the thermal changes described in the introduction, are in fact not included in the Hodgkin-Huxley model as it is presently understood. One has to distinguish between the HH-model from 1952 \cite{Hodgkin1952b} (or modern adaptations of it as in \cite{Howells2012}) that is written in a precise mathematical language and a broader electrochemical viewpoint loosely referring to the function of ion channels. The Hodgkin-Huxley model itself does not explicitly contain either mechanical changes or temperature changes. The only thermodynamic term in the HH equation is the charging of a capacitor with constant capacitance. The latter ad hoc assumption excludes the possibility of mechanical changes in the HH-model, because a thickness change of the membrane will by necessity lead to changes in its capacitance. Therefore, the HH-model cannot be complete. An electrochemical theory that quantitatively contains the mechanical changes and the temperature changes does not yet exist. The implications of the mechanical and thermal findings are discussed in depth in Appendices A and B. 

The reversible heat changes discussed in the introduction are not consistent with the Hodgkin-Huxley model either because 1.\ the magnitude of the heat change is too large (by at least a factor of two ) and 2.\ because a heat change at the site of the membrane will only be positive upon discharge of the membrane capacitor, but not negative upon charging. This was pointed out by Howarth et al. \cite{Howarth1968}. Heat changes are discussed in more detail in Appendix B.

In his textbook from 1964 Alan L. Hodgkin recommended: \textit{"In thinking about the physical basis of the action potential perhaps the most important thing to do at the present moment is to consider whether there are any unexplained observations which have been neglected in an attempt to make the experiments fit into a tidy pattern"} \cite{Hodgkin1964}. Hodgkin was especially concerned about the temperature changes measured during the action potential \cite{Abbott1958}. While Hodgkin's concerns reflect good scientific praxis, neither the measurement of heat nor mechanical changes associated with the action potential have received the attention that they merit.  Here, we have demonstrated that mechanical signals in individual axons from lobster connectives propagate in phase with electrical signals. This is in agreement with very early recordings in squid by Iwasa and Tasaki \cite{Iwasa1980a, Iwasa1980b}. The vertical amplitudes vary from 0.2-1.2 nm. We have also shown that in our nerve preparation pulses traveling in opposite directions pass through each other without significant distortion.  This is expected for electromechanical pulses or solitons. It is, however, unexpected within the framework of a Hodgkin-Huxley mechanism (see also \cite{GonzalezPerez2014}). There exists one old (somewhat unclear) paper from 1949 \cite{Tasaki1949} that reported pulse annihilation in single axons from the sciatic nerve of toads. To our knowledge, this is the only other study besides \cite{GonzalezPerez2014} and the present study investigating pulse collisions in a single axon. Since we report on a different nerve, we cannot claim that this is not possible. However, it would be interesting to study the origin of the deviating result in different preparations. While our findings do not prove any particular mechanism for the action potentials, they do suggest that it is advisable to study changes in thermodynamic variables in addition to voltage and current.

Recently, there have been various reports, both theoretical and experimental, regarding the possibility of mechanical pulse propagation in artificial systems close to transitions and in nerves \cite{Kaufmann1989e, Heimburg2005c, Heimburg2007b, Lautrup2011, Griesbauer2012a, Griesbauer2012b, Shrivastava2014a, Shrivastava2014b}. Heimburg and Jackson \cite{Heimburg2005c} argued that, close to the phase transitions found in biological tissue, electromechanical solitons with properties similar to those of the action potential can travel along the nerve axons.  El Hady and Machta  \cite{ElHady2015} recently argued that, in general, the change of charge on the membrane capacitor leads to electrostrictive forces that must alter the membrane dimensions (see also \cite{Heimburg2012}). In \cite{Heimburg2007c} it was shown that such an electromechanical picture can provide important insights regarding the mechanism of general anesthesia, which is then seen as the familiar physical chemical effect of anesthetics on melting transitions in biomembranes.  It seems likely that the neglect of non-electrical effects has been motivated by the relatively narrow and exclusively electrical focus of the accepted framework provided by Hodgkin and Huxley \cite{Hodgkin1952b}.  The study of the mechanical nature of the nerve pulse promises important insights into the underlying thermodynamic nature of the action potential and its control by thermodynamic variables distinct from voltage.\\

%----- end DISCUSSION --------------------------------------------------------

%----- begin CONCLUSIONS ---------------------------------------
%\section*{Conclusions}

%----- end CONCLUSIONS ---------------------------------------

%----- begin ACKNOWLEDGEMENTS ---------------------------------------
\noindent\textbf{Acknowledgments:} 
This work was supported by the Villum Foundation (VKR 022130).
%----- end ACKNOWLEDGEMENTS -----------------------------------------
%Begin insert on heat in Hodgkin Huxley
\appendix
\section*{Appendix}
\subsection*{Changes in membrane thickness in the HH-\newline model and the soliton theory}

Any voltage applied across a membrane will change the membrane dimensions \cite{Heimburg2012, Mosgaard2015, ElHady2015}. Furthermore, any dimensional change of the membrane will change its capacitance \cite{Heimburg2012}. This effect is called piezoelectricity or electrostriction.  It is independent of the origin of the voltage change. Thus, voltage changes introduced by currents through a protein as assumed in the Hodgkin-Huxley model can also induce thickness changes. This statement, however, does not imply that such mechanical changes are contained in the HH-model or even consistent with it.
%However, the Hodgkin-Huxley model considers the membrane capacitance as constant and thereby rules out that thickness changes of the membrane take place. 
The differential equation describing pulse propagation in the Hodgkin-Huxley model is given by
\begin{equation}
\label{HH1}
\frac{a}{2 R_i}\frac{\partial^2 V}{\partial x^2}=C_m\frac{\partial V}{\partial t}+\sum_i g_i (V-E_i) \;.
\end{equation}
Here, $a$ is the radius of the axon, $R_i$ is the specific resistance of the internal medium, the $g_i$ are the conductances of the membrane for ion species $i$ and the $E_i$ are the corresponding Nernst potentials. $I_c=C_m (dV/dt)$ is the capacitive current assuming constant capacitance $C_m$. However, if membrane dimensions can change, the capacitive current is correctly given by \cite{Heimburg2012}
\begin{equation}
\label{HH2}
I_c=C_m \frac{\partial V}{\partial t}+ V \frac{\partial C_m}{dt} \;.
\end{equation}
Only the second term on the right hand side allows for dimensional changes of the membrane. One might hope that this term was small and could be neglected. However, as shown in \cite{Heimburg2012}, a change in membrane thickness such as the one observed here leads to a change in capacitance of $\approx$50\%. A quick back-of-the-envelope estimate suggests that the two terms of eq. (\ref{HH2}) have a roughly equal order of magnitude. This means that the second term cannot be neglected. Thus, the HH model is at least incomplete. It contains neither the possibility of changes in capacitance, nor electrostrictive forces, nor any other work term not contained in the charging of a capacitor with constant capacitance. If capacitance changes were allowed, the HH equations would assume a different form. In particular, the second term of eq. (\ref{HH2}) and the elastic constants of the membrane would have to be introduced. This requires a thermodynamic theory for the changes in capacitance. This is clearly not contained in the present understanding of the HH-model, and this problem is not easily fixed without introducing detailed knowledge about the thermodynamics of the membrane. In contrast, the soliton model takes precisely such matters into account since it is a hydrodynamic theory based on the changes of the elastic constants and dimensional changes of the membrane during the nervous impulse.  The soliton model therefore provides a thermodynamic basis for the changes in capacitance and for piezoelectricity. Since the membrane capacitance changes proportionally with the membrane thickness, voltage changes will be proportional to the membrane thickness. This is the case for the data reported here, but not for the data by Iwasa \& Tasaki \cite{Iwasa1980a, Iwasa1980b}.

Throughout our text we refer to eq. (\ref{HH1}) when we discuss the Hodgkin-Huxley model. It represents a very specific mindset with well-defined mathematics. There exist modern adaptations for more complicated scenarios \cite{Howells2012} that also assume constant capacitance. We do not refer to more general pictures of the nerve membrane somehow containing ion channels that are not associated to a clearly worked-out mathematical form.

\subsection*{Heat production in nerves}

One of the striking properties of the nerve pulse is the observation of reversible heat production during the action potential. A first phase of heat release is followed by a second phase of heat reabsorption of nearly equal magnitude. We discuss below how heat exchange arises in the Hodgkin-Huxley model and in the soliton model. We further argue that the heat changes are inconsistent with the HH model but consistent with a view where the action potential consists of an adiabatic pulse in the membrane.\\
Hodgkin and Huxley \cite{Hodgkin1952b} modelled the nerve by equivalent circuits containing resistors (the ion channels), capacitors (the membrane) and a battery (the concentration differences of Na$^+$ and K$^+$ ions between inside and outside). In such an electrical view, the resistors heat up when a current flows independent of the direction of the currents (Joule heating). There is no mechanism to cool the membrane.  Looking at thermodynamic analogues of the HH-model, the charging and discharging of a capacitor is caused by the flow of ions from on side to the other, which can be considered as ideal gases that expand through semipermeable walls. The expansion of an ideal gas does not lead to a change in the energy of the ions. Therefore, the work W done by the ions on the capacitor is equivalent to the heat Q that is absorbed from the reservoir (W=-Q).  Thus, the charging of a capacitor leads to the absorption of heat from the reservoir that is exactly equal to the free energy on the capacitor $\frac{1}{2}C_m (V-V_0)^2$ \cite{Mosgaard2015}. Upon discharging the capacitor, this heat is released locally at the site of the membrane. Thus, as expected for an ideal gas, when reversible work is done on a nerve, there is no net heat production after the pulse. This scenario, called the condenser theory was discussed in depth by \cite{Howarth1968}. They made a detailed summary of their findings and concluded amongst others: \textit{"The condenser theory, according to which the positive heat represents the dissipation of electrical energy stored in the membrane capacity, while the negative heat results from the recharging of the capacity, appears unable to account for more than half of the observed temperature changes."} This implies that the observed heat changes are simply too large to be consistent with the charging of the membrane alone. In the condenser theory, the heat is absorbed in the bulk (i.e., the battery) but released locally. Therefore, the condenser theory is not isentropic, and the membrane itself is not adiabatic on the time scale of the nerve pulse. A thermocouple placed directly on the membrane would see an increase of the temperature upon discharge of the capacitor but no decrease in temperature upon charging it as a consequence of ion flows.  It is essential in these thermodynamic analogies to the HH-model that the changes of ion-concentrations in the bulk provide the energy for charging the membrane, and not some energy of the membrane itself. It should also be noted, that the thermodynamic reinterpretation of the HH-model made above is not identical to the HH-model because the latter is based on equivalent circuits. I.e., it is a purely electrical theory.

Howarth et al. conclude that \textit{"It seems probable that the greater part of the initial heat results from changes in the entropy of the nerve membrane when it is depolarized and repolarized".} This is in fact the view of the soliton model. It requires a density change of the membrane from a liquid to a solid state of the membrane. During the first phase of the pulse, the latent heat of the transition is released into the environment of the nerve membrane.  During the second phase, the membrane returns to the liquid state, and the latent heat is reabsorbed. This is exactly the chain of events seen in experimental heat recordings. In \cite{Heimburg2005c}, the transient heat released during the action potential was explained be the energy change of the membrane during a compression. It was found that the heat release is qualitatively and quantitatively consistent with the heat recordings by \cite{Abbott1958, Howarth1968, Howarth1975, Ritchie1985}.
In the soliton model, the reversible heat release is due to the work necessary to compress a charged membrane. Thus, it contains both the charging of the capacitor and the mechanical work performed in order to change the membrane density. Since the physiological membrane is in its liquid state slightly above a melting in the membrane, the soliton model requires that local cooling of the membrane from physiological temperature is able to trigger a nerve pulse, while heating inhibits the nerve pulse. This was in fact found by \cite{Kobatake1971}. The authors argued that this provides evidence in favor of a phase transition in the membrane. In contrast, a more recent study reported that heating by infrared pulses can trigger nerve activity \cite{Shapiro2012}. The authors explained this effect rather by heating of the electrolyte (rather than of the membrane) and a subsequent change in the electrostatics of the electric double layer. They suggested that the heating of the buffer generates a voltage change triggering a pulse. Thus, these two studies may well address different phenomena.
\\

\small{
%\bibliography{literature050409bibdesk}
%\bibliography{/Users/thomasheimburg/Documents/0_Thomas/TH_MSBerichteAntr/Manuskripte/Bibtex_central/Bibdesk_central_literature}
%\bibliography{Bibdesk_central_literature}

\begin{thebibliography}{10}

\bibitem{Goodman1974}
Goodman, C.
\newblock 1974.
\newblock Anatomy of locust ocellar interneurons - constancy and variability.
\newblock J.\ Comp.\ Physiol. 95:185--201.

\bibitem{Hodgkin1952b}
Hodgkin, A.~L., and A.~F. Huxley.
\newblock 1952.
\newblock A quantitative description of membrane current and its application to
  conduction and excitation in nerve.
\newblock J.\ Physiol.\ London 117:500--544.

\bibitem{Iwasa1980a}
Iwasa, K., and I.~Tasaki.
\newblock 1980.
\newblock Mechanical changes in squid giant-axons associated with production of
  action potentials.
\newblock Biochem.\ Biophys.\ Research Comm. 95:1328--1331.

\bibitem{Iwasa1980b}
Iwasa, K., I.~Tasaki, and R.~C. Gibbons.
\newblock 1980.
\newblock Swelling of nerve fibres associated with action potentials.
\newblock Science 210:338--339.

\bibitem{Tasaki1989}
Tasaki, I., K.~Kusano, and M.~Byrne.
\newblock 1989.
\newblock Rapid mechanical and thermal changes in the garfish olfactory nerve
  associated with a propagated impulse.
\newblock Biophys.\ J. 55:1033--1040.

\bibitem{Kim2007}
Kim, G.~H., P.~Kosterin, A.~Obaid, and B.~M. Salzberg.
\newblock 2007.
\newblock A mechanical spike accompanies the action potential in mammalian
  nerve terminals.
\newblock Biophys.\ J. 92:3122--3129.

\bibitem{Wilke1912b}
Wilke, E., and E.~Atzler.
\newblock 1912.
\newblock Experimentelle {B}eitr{\"a}ge zum {P}roblem der {R}eizleitung im
  {N}erven.
\newblock Pfl\"ugers Arch. 146:430--446.

\bibitem{Abbott1958}
Abbott, B.~C., A.~V. Hill, and J.~V. Howarth.
\newblock 1958.
\newblock The positive and negative heat production associated with a nerve
  impulse.
\newblock Proc.\ Roy.\ Soc.\ Lond.\ B 148:149--187.

\bibitem{Howarth1968}
Howarth, J.~V., R.~Keynes, and J.~M. Ritchie.
\newblock 1968.
\newblock The origin of the initial heat associated with a single impulse in
  mammalian non-myelinated nerve fibres.
\newblock J.\ Physiol. 194:745--793.

\bibitem{Howarth1975}
Howarth, J.
\newblock 1975.
\newblock Heat production in non-myelinated nerves.
\newblock Phil. Trans. Royal Soc. Lond. 270:425--432.

\bibitem{Ritchie1985}
Ritchie, J.~M., and R.~D. Keynes.
\newblock 1985.
\newblock The production and absorption of heat associated with electrical
  activity in nerve and electric organ.
\newblock Quart.\ Rev.\ Biophys. 18:451--476.

\bibitem{Hill1912}
Hill, A.~V.
\newblock 1912.
\newblock The absence of temperature changes during the transmission of a
  nervous impulse.
\newblock J.\ Physiol.\ London 43:433--440.

\bibitem{Hodgkin1964}
Hodgkin, A.~L., 1964.
\newblock The conduction of the nervous impulse.
\newblock Liverpool University Press, Liverpool, UK.

\bibitem{Heimburg2005c}
Heimburg, T., and A.~D. Jackson.
\newblock 2005.
\newblock On soliton propagation in biomembranes and nerves.
\newblock Proc.\ Natl.\ Acad.\ Sci.\ USA 102:9790--9795.

\bibitem{Heimburg2007a}
Heimburg, T., 2007.
\newblock Thermal biophysics of membranes.
\newblock Wiley VCH, Berlin, Germany.

\bibitem{Lautrup2011}
Lautrup, B., R.~Appali, A.~D. Jackson, and T.~Heimburg.
\newblock 2011.
\newblock The stability of solitons in biomembranes and nerves.
\newblock Eur.\ Phys.\ J.\ E 34:57.

\bibitem{Heimburg2007b}
Heimburg, T., and A.~D. Jackson.
\newblock 2007.
\newblock On the action potential as a propagating density pulse and the role
  of anesthetics.
\newblock Biophys. Rev. Lett. 2:57--78.

\bibitem{Heimburg2012}
Heimburg, T.
\newblock 2012.
\newblock The capacitance and electromechanical coupling of lipid membranes
  close to transitions. the effect of electrostriction.
\newblock Biophys.\ J. 103:918--929.

\bibitem{GonzalezPerez2014}
{Gonzalez-Perez}, A., R.~Budvytyte, L.~D. Mosgaard, S.~Nissen, and T.~Heimburg.
\newblock 2014.
\newblock Penetration of action potentials during collision in the median and
  lateral giant axons of invertebrates.
\newblock Phys.\ Rev.\ X 4:031047.

\bibitem{Goldman1964}
Goldman, L.
\newblock 1964.
\newblock The effects of stretch on cable and spike parameters of single nerve
  fibres; some implications for the theory of impulse.
\newblock J.\ Physiol.\ London 175:425--444.

\bibitem{Griesbauer2012a}
Griesbauer, J., S.~B\"ossinger, A.~Wixforth, and M.~F. Schneider.
\newblock 2012.
\newblock Propagation of 2d pressure pulses in lipid monolayers and its
  possible implications for biology.
\newblock Phys.\ Rev.\ Lett. 108:198103.

\bibitem{Griesbauer2012b}
Griesbauer, J., S.~B\"ossinger, A.~Wixforth, and M.~F. Schneider.
\newblock 2012.
\newblock Simultaneously propagating voltage and pressure pulses in lipid
  monolayers of pork brain and synthetic lipids.
\newblock Phys.\ Rev.\ E 86:061909.

\bibitem{Evans1976}
Evans, P.~D., E.~A. Kravitz, B.~R. Talamo, and B.~G. Wallace.
\newblock 1976.
\newblock The association of octopamine with specific neurones along lobster
  nerve trunks.
\newblock The Journal of physiology 262:51--70.

\bibitem{Wright1958}
Wright, E.~B., and J.~P. Reuben.
\newblock 1958.
\newblock A comparative study of some excitability properties of the giant
  axons of the ventral nerve cord of the lobster, including the recovery of
  excitability following an impulse.
\newblock J.\ Cell\ Comp.\ Physiol. 51:13--28.

\bibitem{ElHady2015}
{El Hady}, A., and B.~B. Machta.
\newblock 2015.
\newblock Mechanical surface waves accompany action potential propagation.
\newblock Nat.\ Communications 6:6697.

\bibitem{Tasaki1949}
Tasaki, I.
\newblock 1949.
\newblock Collision of two nerve impulses in the nerve fiber.
\newblock Biochim.\ Biophys.\ Acta 3:494--497.

\bibitem{Salzberg1989}
Salzberg, B.~M.
\newblock 1989.
\newblock Optical recording of voltage changes in nerve terminals and in fine
  neuronal processes.
\newblock Ann.\ Rev.\ Physiol. 51:507--526.

\bibitem{Zhang2001}
Zhang, P.-C., A.~M. Keleshian, and F.~Sachs.
\newblock 2001.
\newblock Voltage-induced membrane movement.
\newblock Nature 413:428--432.

\bibitem{Nguyen2012}
Nguyen, T.~D., N.~Deshmukh, J.~M. Nagarah, T.~Kramer, P.~K. Purohit, M.~J.
  Berry, and M.~C. {McAlpine}.
\newblock 2012.
\newblock Piezoelectric nanoribbons for monitoring cellular deformations.
\newblock Nat.\ Nanotechnol. 7:587--593.

\bibitem{Howells2012}
Howells, J., L.~Trevillion, H.~Bostock, and D.~Burke.
\newblock 2012.
\newblock The voltage dependence of {I}$_h$ in human myelinated axons.
\newblock J.\ Physiol. 590:1625--1640.

\bibitem{Kaufmann1989e}
Kaufmann, K., 1989.
\newblock Lipid membrane.
\newblock http://membranes.nbi.dk/Kaufmann/pdf/\-Kaufmann\_book5\_org.pdf,
  Caruaru.

\bibitem{Shrivastava2014a}
Shrivastava, S., and M.~F. Schneider.
\newblock 2014.
\newblock Evidence for two-dimensional solitary sound waves in a lipid
  controlled interface and its implications for biological signalling.
\newblock J.\ R.\ Soc.\ Interface 11:20140098.

\bibitem{Shrivastava2014b}
Shrivastava, S., K.~H. Kang, and M.~F. Schneider.
\newblock 2015.
\newblock Solitary shock waves and adiabatic phase transition in lipid
  interfaces and nerves.
\newblock Phys.\ Rev.\ E 91:012715.

\bibitem{Heimburg2007c}
Heimburg, T., and A.~D. Jackson.
\newblock 2007.
\newblock The thermodynamics of general anesthesia.
\newblock Biophys.\ J. 92:3159--3165.

\bibitem{Mosgaard2015}
Mosgaard, L.~D., K.~Zecchi, and T.~Heimburg.
\newblock 2015.
\newblock Mechano-capacitive properties of polarized membranes.
\newblock Soft Matter DOI: 10.1039/C5SM01519G.

\bibitem{Kobatake1971}
Kobatake, Y., I.~Tasaki, and A.~Watanabe.
\newblock 1971.
\newblock Phase transition in membrane with reference to nerve excitation.
\newblock Adv.\ Biophys. 208:1--31.

\bibitem{Shapiro2012}
Shapiro, M.~G., K.~Homma, S.~Villareal, C.-P. Richter, and F.~Bezanilla.
\newblock 2012.
\newblock Infrared light excites cells by changing their electrical
  capacitance.
\newblock Nat.\ Communications 3:736.

\end{thebibliography}
%\bibliographystyle{biophysj2005}
}

%====================================================

\footnotesize{

}
\normalsize
\end{document}